\documentclass[12pt]{article}
\input epsf

\begin{document}

\title{Quark distribution functions in
the chiral quark-soliton model: cancellation of
quantum anomalies}

\author{K. Goeke$^a$, P.V. Pobylitsa$^{a,b}$, M.V. Polyakov$^{a,b}$, \\
P. Schweitzer$^a$, D. Urbano$^{a,c}$\\
\footnotesize\it $^a$ Institute for Theoretical Physics II, Ruhr University
Bochum, Germany\\
\footnotesize\it $^b$ Petersburg Nuclear Physics Institute, Gatchina, St. Petersburg 188350,
Russia\\
\footnotesize\it $^c$ Faculdade de Engenharia da Universidade do Porto, 4000
 Porto, Portugal}
\date{}
\maketitle

\vspace{-9cm}
\begin{flushright}
RUB/TPII-18/99
\end{flushright}
\vspace{7cm}
\begin{abstract}
\noindent In the framework of the chiral quark-soliton model of the nucleon
we investigate the properties of the polarized quark distribution. In
particular we analyse the so called anomalous difference between the
representations for the quark distribution functions in terms of occupied
and non-occupied quark states. By an explicit analytical calculation it is
shown that this anomaly is absent in the polarized isoscalar distribution 
$\Delta u + \Delta d$, which is ultraviolet finite. In the case of the
polarized isovector quark distribution $\Delta u-\Delta d$ the anomaly can be
cancelled by a Pauli-Villars subtraction which is also needed for the
regularization of the ultraviolet divergence.
\end{abstract}

\section{Introduction}

Recently a rather successful program of computing the quark distribution
functions in the framework of the effective quark-soliton model was developed \cite
{DPPPW96,DPPPW97,PP96,PPGWW,WK99,WGR-97}. The quark soliton model
\cite{DPP} includes
the chiral pion field $U=e^{i\pi ^{a}\tau ^{a}/F_{\pi }}$ and the quark field
$\psi $ whose interaction is described by the Lagrangian
\begin{equation}
L=\bar{\psi}(i\gamma ^{\mu }\partial _{\mu }-
MU^{\gamma _{5}})\psi \, .
\end{equation}
In the mean field approximation (justified in the limit of the large number
of quark colors $N_{c}$ \cite{Witten}) the nucleon arises as a soliton of
the chiral field $U$
\begin{equation}
U(x)=\exp [i(n^{a}\tau ^{a})P(r)],\quad n^{a}=\frac{x^{a}}{r},\quad r=|x| \, .
\label{hedgehog}
\end{equation}

This effective theory allows a quantum field-theoretical approach to the
calculation of the quark and antiquark distributions in the nucleon. In
contrast to naive quark composite models and to the bag model here we have a
consistent approach reproducing the main features of the QCD parton model
like positivity of the quark and antiquark distributions, various sum rules
etc.

In terms of the quark degrees of freedom this picture of the nucleon
corresponds to occupying with $N_{c}=3$ quarks the negative continuum levels as well as the valence level
of the one-particle Dirac Hamiltonian $H$%
\begin{equation}
H=-i\gamma ^{0}\gamma ^{k}\partial _{k}+M\gamma ^{0}U^{\gamma _{5}} \, ,
\label{H-Dirac}
\end{equation}
in the background soliton field $U$. For the pion field (\ref{hedgehog})
one can find the spectrum of the Hamiltonian (\ref{H-Dirac})
\begin{equation}
H|n\rangle = E_n |n\rangle \, .
\label{eigens}
\end{equation}

Various nucleon observables can be naturally represented as sums over
eigenstates $|n\rangle $ of the Dirac Hamiltonian $H$. For example, the
nucleon mass $M_{N}$ is given by the sum over all occupied states or
alternatively by the minus sum over all non-occupied states
\begin{equation}
M_{N}=\sum\limits_{n,occ}E_{n} =-\sum\limits_{n,non-occ} E_{n} \label{M-N-sum} \, .
\end{equation}
In this expression the subtraction of similar sums is implied
where the eigenstates $|n\rangle$ and the eigenvalues $E_n$ are
replaced by those of the free Hamiltonian,
\begin{equation}
H_0 |n^{(0)}\rangle = E^{(0)}_n |n^{(0)}\rangle ,\quad \, H_{0}=-i\gamma
^{0}\gamma ^{k}\partial _{k}+M\gamma ^{0} .\quad \,
\label{freeham}
\end{equation}

The physical reason for the existence of the two equivalent expressions in
(\ref{M-N-sum}) is
that the polarized Dirac sea picture can be formulated either in terms of
quark or in terms of antiquark states (occupied antiquark states correspond
to non-occupied quark states).

Formally the equivalence of two representations (\ref{M-N-sum}) for $M_{N}$
follows from the identity
\begin{equation}
\sum\limits_{n,occ} E_{n} +\sum\limits_{n,non-occ} E_{n} =\sum\limits_{n}E_{n}=Tr%
\,H=0  \, .\label{trace-zero}
\end{equation}
At the last step we took into account that the trace of $H$ over the spin
indices vanishes. Strictly speaking, this naive argument is not safe since
the sums (\ref{M-N-sum}) over the occupied and non-occupied states are
ultraviolet divergent and must be regularized. In principle, the ultraviolet
regularization could lead to an anomalous difference between the summation
over occupied and non-occupied states but in the case of the nucleon mass (%
\ref{M-N-sum}) one can check that in the regularizations like Pauli-Villars
or proper-time ones the anomaly is absent. E.g. the proper-time regularized
version of (\ref{trace-zero}) with the ultraviolet cutoff $\Lambda $ is
\begin{equation}
\lim_{\Lambda \rightarrow \infty }\sum\limits_{n}E_{n}\exp \left[
-E_{n}^{2}/\Lambda ^{2}\right] =0 \, .
\end{equation}

We can reformulate this verbally as the ``absence of the anomaly'' in the
nucleon mass $M_{N}$ (in the proper time regularization). The usage of the
word ``anomaly'' is invoked by  the similarity with the axial anomaly which
can be interpreted as nonvanishing trace of $\gamma _{5}$ in e.g. the
proper-time regularization.

The main object of interest in this paper is the study of the quark
distribution functions. In the mean field approach (justified in the large $%
N_{c}$ limit) the quark distributions can be represented as single or double
sums over occupied or non-occupied one-particle eigenstates (\ref{eigens}) of the Dirac
Hamiltonian (\ref{H-Dirac}). We shall see that for the same parton distribution one
can
write two naively equivalent representations but whether this
equivalence persists or not when one takes into account the ultraviolet
regularization is a rather subtle question and the situation is different for different
distributions. Moreover, even in the limit of the large cutoff, the cancellation of this
 anomalous difference between
the naively equivalent representations is sensitive to the regularization used.
\vspace{0.1cm}

Let us start from the unpolarized isosinglet quark distribution $u(x)+d(x)$
which is given by the following expressions (\cite{DPPPW96}) in the leading
order of the $1/N_{c}$ expansion
\[
u(x)+d(x)=N_{c}\sum\limits_{n,occ}\int \frac{d^{3}p}{(2\pi )^{3}}\delta
\left( \frac{p^{3}+E_{n}}{M_{N}}-x\right) \langle n|\mathbf{p}\rangle
(1+\gamma ^{0}\gamma ^{3})\langle \mathbf{p}|n\rangle
\]
\begin{equation}
\quad \quad \quad =-N_{c}\sum\limits_{n,non-occ}\int \frac{d^{3}p}{(2\pi )^{3}%
}\delta \left( \frac{p^{3}+E_{n}}{M_{N}}-x\right) \langle n|\mathbf{p}%
\rangle (1+\gamma ^{0}\gamma ^{3})\langle \mathbf{p}|n\rangle
\label{u-plus-d} \, .
\end{equation}
Also here the subtraction of similar sums  with the eigenstates and
eingenvalues of the Hamiltonian (\ref{H-Dirac}) replaced by those of the free
Hamiltonian (\ref{freeham})  is implied.
The result (\ref{u-plus-d}) has a transparent physical meaning of the
probability to find a
quark with momentum fraction $x$ in the nucleon in the infinite momentum
frame. In ref. \cite{DPPPW96} it was shown that in the Pauli-Villars
regularization the sums over occupied and non-occupied states in (\ref
{u-plus-d}) really give the same result.

We stress that the fact of the equivalence of the two representations for
parton distributions is crucial for the positivity of unpolarized
distributions and for the validity of various sum rules inherited by the
model from QCD \cite{DPPPW96}. Therefore the check of this equivalence is an
essential part of the calculation of parton distributions in the chiral
soliton model.

Now let us turn to the polarized quark distributions. In the leading order
of the $1/N_c$ expansion only the isovector polarized distribution survives
\[
\Delta u(x)-\Delta d(x)=-\frac 13N_c\sum\limits_{n,occ}\int \frac{d^3p}{%
(2\pi )^3}\delta \left( \frac{p^3+E_n}{M_N}-x\right)
\]
\begin{equation}
\times \langle n|\mathbf{p}\rangle (1+\gamma ^0\gamma ^3)\tau ^3\gamma
^5\langle \mathbf{p}|n\rangle  \, . \label{Du-minus-Dd-occ}
\end{equation}
Compared to the expression (\ref{u-plus-d}) for $u(x)+d(x)$ here we have an
extra factor $\tau ^3\gamma ^5$ which reflects the fact that now we deal
with the isovector polarized distribution. The factor of $1/3$ comes from a
careful treatment of the rotation of the soliton \cite{DPPPW96}.

One can ask whether the summation over the occupied quark states in (\ref
{Du-minus-Dd-occ}) can be replaced by the summation over non-occupied states
\[
\Delta u(x)-\Delta d(x)=\frac 13N_c\sum\limits_{n,non-occ}\int \frac{d^3p}{%
(2\pi )^3}\delta \left( \frac{p^3+E_n}{M_N}-x\right)
\]
\begin{equation}
\times \langle n|\mathbf{p}\rangle (1+\gamma ^0\gamma ^3)\tau ^3\gamma
^5\langle \mathbf{p}|n\rangle  \, . \label{Du-minus-Dd-non-occ}
\end{equation}
In this paper we shall show that in the case of the Pauli-Villars
regularization (the sum over states $n$ in (\ref{Du-minus-Dd-occ}) is
logarithmically divergent) the two representations (\ref{Du-minus-Dd-occ})
and (\ref{Du-minus-Dd-non-occ}) are really equivalent.

We stress that the equivalence of the summation over the occupied and
non-occupied states is very sensitive to the choice of the regularization.
For example, if instead of the Pauli--Villars regularization we simply cut
the summation over quark states in (\ref{Du-minus-Dd-occ}) including only
states with $|E_{n}|<\omega _{0}$ then a nonzero difference between the two
representations (\ref{Du-minus-Dd-occ}) and (\ref{Du-minus-Dd-non-occ}) will
remain even in the limit of the infinite cutoff $\omega _{0}\rightarrow
\infty $. The mechanism how this anomalous difference appears is similar in
many respects to the famous axial anomaly. In particular, such similarity
manifests itself in the fact that the anomalous difference between the two
representations (\ref{Du-minus-Dd-occ}) and (\ref{Du-minus-Dd-non-occ}) can
be computed analytically in the limit $\omega _{0}\rightarrow \infty $. The
calculation of the anomalous difference is presented in this paper.

Although the regularization including only states with $|E_{n}|<\omega _{0}$
is not acceptable as a physical one and the Pauli--Villars regularization is
more preferable in this respect, we want to emphasize that in the practical
calculations based on the numerical diagonalization of the Dirac operator in
the background soliton field, the $|E_{n}|<\omega _{0}$ regularization
appears naturally. Indeed, in the numerical calculation one can work only
with a finite amount of quark states so that one actually uses both
Pauli-Villars subtraction (with the regulator mass $M_{PV}$) and the $%
|E_{n}|<\omega _{0}$ regularization. The pure Pauli-Villars subtraction is
simulated by working with $\omega _{0}\gg M_{PV}$. The numerical calculation
is rather involved and the analytical result for the anomaly in the $%
|E_{n}|<\omega _{0}$ regularization is very helpful for the control of
numerics even if the anomaly cancels after the Pauli-Villars subtraction.

Now let us turn to the polarized isoscalar quark distribution $\Delta
u(x)+\Delta d(x)$ which gets the first nonzero contribution only in the
subleading order of the $1/N_{c}$ expansion
\[
\Delta u(x)+\Delta d(x)=\frac{N_{c}M_{N}}{2I}\sum\limits_{m,all}\sum%
\limits_{n,occ}\frac{1}{E_{n}-E_{m}}
\]
\[
\times \langle n|\tau ^{3}|m\rangle \langle m|(1+\gamma ^{0}\gamma
^{3})\gamma ^{5}\delta (E_{n}+P^{3}-xM_{N})|n\rangle
\]
\begin{equation}
+\frac{N_{c}}{4I}\,\frac{\partial }{\partial x}\sum\limits_{n,occ}\langle
n|(1+\gamma ^{0}\gamma ^{3})\tau ^{3}\gamma ^{5}\delta
(E_{n}+P^{3}-xM_{N})|n\rangle  \, . \label{Du-plus-Dd-occ}
\end{equation}
Here $P^{3}$ is the quark momentum projection on the third axis
\begin{equation}
P^{3}=-i\frac{\partial }{\partial x^{3}}  \, , \label{P-3-def}
\end{equation}
and $I$ is the moment of inertia of the soliton \cite{Review}.

Another representation for $\Delta u+\Delta d$ can be written in terms of
the summation over non-occupied states $n$

\[
\Delta u(x)+\Delta d(x)=-\frac{N_{c}M_{N}}{2I}\sum\limits_{m,all}\sum%
\limits_{n,non-occ}\frac{1}{E_{n}-E_{m}}
\]
\[
\times \langle n|\tau ^{3}|m\rangle \langle m|(1+\gamma ^{0}\gamma
^{3})\gamma ^{5}\delta (E_{n}+P^{3}-xM_{N})|n\rangle
\]
\begin{equation}
-\frac{N_{c}}{4I}\,\frac{\partial }{\partial x}\sum\limits_{n,non-occ}\langle
n|(1+\gamma ^{0}\gamma ^{3})\tau ^{3}\gamma ^{5}\delta
(E_{n}+P^{3}-xM_{N})|n\rangle  \, . \label{Du-plus-Dd-non-occ}
\end{equation}
The numerical calculation of $\Delta u+\Delta d$ with the Pauli--Villars
subtraction was presented in paper \cite{WK99}. Unfortunately there the question
about the equivalence of the two representation (\ref{Du-plus-Dd-occ}) and (%
\ref{Du-plus-Dd-non-occ}) was not investigated properly.
Also the Pauli-Villars subtraction was used in paper \cite{WK99} without
proper justification.

In this paper we show that if one cuts the sum over occupied (non-occupied)
states $n$ allowing only $|E_n|<\omega _0$ in the eqs. (\ref{Du-plus-Dd-occ}%
), (\ref{Du-plus-Dd-non-occ}) then in the infinite cutoff limit $\omega
_0\rightarrow \infty $

1) both representations (\ref{Du-plus-Dd-occ}) and (\ref{Du-plus-Dd-non-occ})
have a finite limit (i.e. $\Delta u(x)+\Delta d(x)$ has no ultraviolet
divergences),

2) the two representations (\ref{Du-plus-Dd-occ}), (\ref{Du-plus-Dd-non-occ})
give the same result.

Comparing the last terms in the rhs of representations (\ref{Du-plus-Dd-occ}%
) and (\ref{Du-plus-Dd-non-occ}) for $\Delta u+\Delta d$ with expressions (%
\ref{Du-minus-Dd-occ}) and (\ref{Du-minus-Dd-non-occ}) for $\Delta u-\Delta d$
we see that the total expression for $\Delta u+\Delta d$ contains a
contribution proportional to $\frac{\partial }{\partial x}\left[ \Delta
u(x)-\Delta d(x)\right] $.

Therefore we start our analysis by investigating the anomaly of $\Delta
u-\Delta d$ which we do in section \ref{du-minus-dd-section}. In section \ref
{du-plus-dd-section} we show by explicit calculation that for the quark
distribution $\Delta u+\Delta d$ there is no anomalous difference between
the summations over occupied and non-occupied states. In section \ref
{num-res-section} we discuss the numerical results and compare them the GRSV
parametrization of experimental data.

\section{Anomaly of $\Delta u(x) - \Delta d(x)$}

\label{du-minus-dd-section}

As it was explained in the introduction one of our aims is to investigate
whether the two representations (\ref{Du-minus-Dd-occ}) and (\ref
{Du-minus-Dd-non-occ}) for the polarized isovector quark distribution $\Delta
u(x)-\Delta d(x)$ are equivalent. The answer to this question is sensitive
to the ultraviolet regularization. Let us start from the regularization that
allows only the quark states $n$ with $|E_n|<\omega _0$. In this
regularization eq. (\ref{Du-minus-Dd-occ}) can be rewritten as follows.
\[
\left[ \Delta u(x)-\Delta d(x)\right] _{occ}^{\omega _0}
\]
\begin{equation}
=-\frac 13N_cM_N\int\limits_{-\omega _0}^{E_{lev}+0}d\omega \mathrm{Tr}\left[
\delta (H-\omega )\delta (\omega +P^3-xM_N)\tau ^3(1+\gamma ^0\gamma
^3)\gamma _5\right]  \, . \label{du-minus-dd-occ}
\end{equation}
Here $H$ is the Dirac Hamiltonian (\ref{H-Dirac}) and $P^3$ is momentum
operator (\ref{P-3-def}). Similarly, representation
(\ref{Du-minus-Dd-non-occ}) becomes
\[
\left[ \Delta u(x)-\Delta d(x)\right] _{non-occ}^{\omega _0}
\]
\begin{equation}
=\frac 13N_cM_N\int\limits_{E_{lev}+0}^{\omega _0}d\omega \mathrm{Tr}\left[
\delta (H-\omega )\delta (\omega +P^3-xM_N)\tau ^3(1+\gamma ^0\gamma
^3)\gamma _5\right]  \, . \label{du-minus-dd-non-occ}
\end{equation}
The main results of this section can be formulated as follows

1) Both $\left[ \Delta u(x)-\Delta d(x)\right] _{occ}^{\omega _0}$ and $%
\left[ \Delta u(x)-\Delta d(x)\right] _{non-occ}^{\omega _0}$ are
logarithmically divergent in the limit of large cutoff $\omega _0\rightarrow
\infty $

\[
\left[ \Delta u(x)-\Delta d(x)\right] _{occ}^{\omega _0}\sim \left[ \Delta
u(x)-\Delta d(x)\right] _{non-occ}^{\omega _0}=\frac{N_cM_NM^2}{12\pi ^2}\ln
\frac{\omega _0}M
\]

\begin{equation}
\times \int \frac{d^3k}{(2\pi )^3}\mathrm{Sp}_{fl}\left[ (\tilde U[\mathbf{k}%
])^{+}\tau ^3\tilde U(\mathbf{k})\right] \theta \left( k^3-|x|M_N\right)
+\ldots  \, . \label{Du-minus-DD-divergence}
\end{equation}

2) In the difference $\left[ \Delta u(x)-\Delta d(x)\right] _{occ}^{\omega
_{0}}-\left[ \Delta u(x)-\Delta d(x)\right] _{non-occ}^{\omega _{0}}$ the
ultraviolet divergences cancel and the $\omega _{0}\rightarrow \infty $
limit of this difference reduces to the following finite expression
\[
\lim_{\omega _{0}\rightarrow \infty }\left\{ \left[ \Delta u(x)-\Delta d(x)%
\right] _{occ}^{\omega _{0}}-\left[ \Delta u(x)-\Delta d(x)\right]
_{non-occ}^{\omega _{0}}\right\}
\]
\begin{equation}
=-\frac{1}{12\pi ^{2}}N_{c}M_{N}M^{2}\int \frac{d^{3}k}{(2\pi )^{3}}\log
\frac{|xM_{N}+k^{3}|}{|xM_{N}|}\mathrm{Sp}_{fl}\left[ \tau ^{3}(\tilde{U}[%
\mathbf{k}])^{+}\tilde{U}(\mathbf{k})\right]  \, .
\label{du-minus-dd-anomaly-res}
\end{equation}
where $\tilde{U}(\mathbf{k})$ is the Fourier transform of the chiral mean
field $U(\mathbf{r})$ entering the Dirac Hamiltonian (\ref{H-Dirac})

\begin{equation}
\tilde U(\mathbf{k})=\int d^3\mathbf{r\,}e^{-i(\mathbf{kr})}\left[ U(\mathbf{%
r})-1\right]  \, . \label{U-Fourier-0}
\end{equation}
Note that $\left[ \Delta u(x)-\Delta d(x)\right] _{occ}^{\omega _0}$ and $%
\left[ \Delta u(x)-\Delta d(x)\right] _{non-occ}^{\omega _0}$ separately are
given by complicated functional traces (\ref{du-minus-dd-occ}) and (\ref
{du-minus-dd-non-occ}) which can be computed only numerically. The fact that
the anomalous difference between the representations in terms of the
occupied and non-occupied states reduces to a simple momentum integral (\ref
{du-minus-dd-anomaly-res}) is highly nontrivial and is similar to the well
known fact that the famous axial anomaly gets its contribution only from the
simplest diagram.

The fact that the divergence (\ref{Du-minus-DD-divergence}) is proportional
to $M^2$ means that this divergence can be removed by the Pauli--Villars
subtraction so that the following combinations are finite
\[
\left[ \Delta u(x)-\Delta d(x)\right] _{occ}^{PV}
\]
\begin{equation}
=\lim_{\omega _0\rightarrow \infty }\left\{ \left[ \Delta u(x)-\Delta d(x)%
\right] _{occ}^{\omega _0,M}-\frac{M^2}{M_{PV}^2}\left[ \Delta u(x)-\Delta
d(x)\right] _{occ}^{\omega _0,M_{PV}}\right\}
\end{equation}
\[
\left[ \Delta u(x)-\Delta d(x)\right] _{non-occ}^{PV}
\]
\begin{equation}
=\lim_{\omega _0\rightarrow \infty }\left\{ \left[ \Delta u(x)-\Delta d(x)%
\right] _{non-occ}^{\omega _0,M}-\frac{M^2}{M_{PV}^2}\left[ \Delta
u(x)-\Delta d(x)\right] _{non-occ}^{\omega _0,M_{PV}}\right\} \, .
\end{equation}
Next since the anomaly (\ref{du-minus-dd-anomaly-res}) is proportional to $%
M^2$ we see that in the Pauli-Villars regularization the summation over
occupied and non-occupied states gives the same results:
\begin{equation}
\left[ \Delta u(x)-\Delta d(x)\right] _{occ}^{PV}=\left[ \Delta u(x)-\Delta
d(x)\right] _{non-occ}^{PV} \, .
\end{equation}

Now let us turn to the derivation of the result (\ref
{du-minus-dd-anomaly-res}) for the anomalous difference between the
summation over occupied and non-occupied states. Subtracting (\ref
{du-minus-dd-non-occ}) from (\ref{du-minus-dd-occ}) we obtain

\[
\left[ \Delta u(x)-\Delta d(x)\right] _{occ}^{\omega _0}-\left[ \Delta
u(x)-\Delta d(x)\right] _{non-occ}^{\omega _0}
\]
\begin{equation}
=-\frac 13N_cM_N\int\limits_{-\omega _0}^{\omega _0}d\omega \mathrm{Tr}\left[
\delta (H-\omega )\delta (\omega +P^3-xM_N)\tau ^3(1+\gamma ^0\gamma
^3)\gamma _5\right]  \, . \label{du-minus-dd-anomaly-start}
\end{equation}
We use the following representation for the operator delta function $\delta
(H-\omega )$%
\begin{equation}
\delta (H-\omega )=\frac{\mathrm{sign}\omega }{2\pi i}\left[ \frac
1{H^2-\omega ^2-i0}-\frac 1{H^2-\omega ^2+i0}\right] (H+\omega ) \, .
\end{equation}
The squared Dirac Hamiltonian (\ref{H-Dirac}) is
\begin{equation}
H^2=-\partial ^2+M^2+iM(\gamma ^k\partial _kU^{\gamma _5}) \, .
\end{equation}
Now (\ref{du-minus-dd-anomaly-start}) takes the form
\[
\left[ \Delta u(x)-\Delta d(x)\right] _{occ}^{\omega _0}-\left[ \Delta
u(x)-\Delta d(x)\right] _{non-occ}^{\omega _0}=-\frac 23N_cM_N\mathrm{Im}%
\int\limits_{-\omega _0}^{\omega _0}\frac{d\omega }{2\pi }\mathrm{sign}%
\omega
\]

\[
\times \mathrm{Tr}\left\{ \frac{1}{-\partial ^{2}+M^{2}-\omega
^{2}-i0+iM(\gamma ^{k}\partial _{k}U^{\gamma _{5}})}(\omega -i\gamma
^{0}\gamma ^{k}\partial _{k}+\gamma ^{0}MU^{\gamma _{5}})\right.
\]
\begin{equation}
\left. \times \delta (\omega +P^{3}-xM_{N})\tau ^{3}(1+\gamma ^{0}\gamma
^{3})\gamma _{5}\right\}  \, . \label{du-m-dd-start-0}
\end{equation}
Next we expand the ``propagator'' in the rhs in powers of $iM(\gamma
^{k}\partial _{k}U^{\gamma _{5}})$

\[
\frac 1{-\partial ^2+M^2-\omega ^2-i0+iM(\gamma ^k\partial _kU^{\gamma
_5})}=\frac 1{-\partial ^2+M^2-\omega ^2-i0}
\]
\begin{equation}
-\frac 1{-\partial ^2+M^2-\omega ^2-i0}iM(\gamma ^k\partial _kU^{\gamma
_5})\frac 1{-\partial ^2+M^2-\omega ^2-i0}+\ldots  \, . \label{dU-expansion}
\end{equation}
The first nonvanishing contribution to (\ref{du-m-dd-start-0}) comes from
the term linear in $iM(\gamma ^k\partial _kU^{\gamma _5})$%
\[
\left[ \Delta u(x)-\Delta d(x)\right] _{occ}^{\omega _0}-\left[ \Delta
u(x)-\Delta d(x)\right] _{non-occ}^{\omega _0}=-\frac 23N_cM_N\mathrm{Im}%
\int\limits_{-\omega _0}^{\omega _0}\frac{d\omega }{2\pi }
\]
\[
\times \mathrm{Tr}\left\{ \frac 1{-\partial ^2+M^2-\omega ^2-i0}
[-iM(\partial _kU^{\gamma _5})]\frac 1{-\partial ^2+M^2-\omega ^2-i0 }
\right.
\]
\begin{equation}
\left. \times (\omega -i\gamma ^0\gamma ^k\partial _k+\gamma ^0MU^{\gamma
_5})(1+\gamma ^0\gamma ^3)\gamma _5\gamma ^k\delta (\omega +P^3-xM_N)\tau
^3\right\} \, .
\end{equation}
Computing the trace over the spin indices and turning to the momentum
representation according to (\ref{U-Fourier-0}) we arrive at
\[
\left[ \Delta u(x)-\Delta d(x)\right] _{occ}^{\omega _0}-\left[ \Delta
u(x)-\Delta d(x)\right] _{non-occ}^{\omega _0}
\]
\[
=\frac 83N_cM_NM^2\mathrm{Im}\int \frac{d^3k}{(2\pi )^3}k^3\mathrm{Sp}%
_{fl}\left\{ (\tilde U[\mathbf{k}])^{+}\tilde U(\mathbf{k})\tau ^3\right\}
\]
\begin{equation}
\times \int\limits_{-\omega _0}^{\omega _0}\frac{d\omega }{2\pi }\int \frac{%
d^3p}{(2\pi )^3}\frac{\mathrm{sign}\omega }{|\mathbf{k}+\mathbf{p}%
|^2+M^2-\omega ^2-i0}\frac{\delta (\omega +p^3-xM_N)}{|\mathbf{p}%
|^2+M^2-\omega ^2-i0} \, .
\end{equation}
The calculation of the integral over $\omega $ and $\mathbf{p}$ is
straightforward and in the limit $\omega _0\rightarrow \infty $ one arrives
at final result (\ref{du-minus-dd-anomaly-res}).

Note that the limit $\omega _0\rightarrow \infty $ should be taken after
computing the $\omega $ and $\mathbf{p}$ integrals. Otherwise if one first computes
integrals $\omega $ and $p^3$ at fixed $p^{\perp }$ and with $\omega
_0=\infty $ then one gets zero
\begin{equation}
\int\limits_{-\infty }^\infty \frac{d\omega }{2\pi }\int\limits_{-\infty
}^\infty \frac{dp^3}{2\pi }\frac{\mathrm{sign}\omega }{|\mathbf{k}+\mathbf{p}%
|^2+M^2-\omega ^2-i0}\frac{\delta (\omega +p^3-xM_N)}{|\mathbf{p}%
|^2+M^2-\omega ^2-i0}=0  \, . \label{omega-p3-zero}
\end{equation}
Actually after integrations over $\omega $ (in the interval $-\omega
_0<\omega <\omega _0$) and $p^3$ the integration over $|\mathbf{p}^{\perp }|$
is restricted at large $\omega _0$ to the interval
\begin{equation}
2\omega _0\min \left[ |xM_N|,|xM_N+k^3|\right] <|\mathbf{p}^{\perp
}|^2<2\omega _0\max \left[ |xM_N|,|xM_N+k^3|\right] \, .
\end{equation}
In the limit of large cutoff $\omega _0\rightarrow \infty $ this interval of
$\mathbf{p}^{\perp }$ is shifted to infinity which explains why the nonzero
result (\ref{du-minus-dd-anomaly-res}) is compatible with the vanishing
integral (\ref{omega-p3-zero}).

A similar phenomenon occurs with the $\mathbf{p}^{\perp }$ integration
region in the contributions coming from the higher terms of the expansion (%
\ref{dU-expansion}). However, since the integrands of these higher order
terms decay faster at large $|\mathbf{p}^{\perp }|$ these higher order terms
give vanishing contribution to the anomalous difference (\ref
{du-minus-dd-anomaly-res}) in the limit of the large cutoff $\omega
_{0}\rightarrow \infty $.

Restricting the integration over $\omega $ in eq. (\ref{du-m-dd-start-0}) to
the interval $-\omega _{0}<\omega <0$ or to $0<\omega <\omega _{0}$ we can
investigate separate distribution functions $\left[ \Delta u(x)-\Delta d(x)%
\right] _{occ}^{\omega _{0}}$ or $\left[ \Delta u(x)-\Delta d(x)\right]
_{non-occ}^{\omega _{0}}$. In this case one gets nonzero contributions from
all terms of the infinite series (\ref{dU-expansion}). However, it is not
difficult to check that only the first nonvanishing term of this expansion
is logarithmically divergent in the limit of large cutoff $\omega
_{0}\rightarrow \infty $ and this logarithmic divergence is given by (\ref
{Du-minus-DD-divergence}). This logarithmic divergence is proportional to $%
M^{2}$ and therefore in our previous calculation of $\Delta u-\Delta d$ we
could regularize it by the Pauli-Villars subtraction. Moreover, since the
anomaly (\ref{du-minus-dd-anomaly-res}) is also proportional to $M^{2}$ it
is cancelled by the same Pauli-Villars subtraction \cite{DPPPW97}.

\section{Cancellation of the anomaly of $\Delta u(x)+\Delta d(x)$}

\label{du-plus-dd-section}

Now we turn to the investigation of $\Delta u(x)+\Delta d(x)$. The $\omega
_0 $ cutoff version of (\ref{Du-plus-Dd-occ}) is
\[
\left[ \Delta u(x)+\Delta d(x)\right] _{occ}^{\omega _0}=\frac{N_cM_N}{2I}%
\sum\limits_m\sum\limits_{-\omega _0<E_n\le E_{lev}}\frac 1{E_n-E_m}
\]
\[
\times \langle n|\tau ^3|m\rangle \langle m|(1+\gamma ^0\gamma ^3)\gamma
^5\delta (E_n+P^3-xM_N)|n\rangle
\]
\begin{equation}
+\frac{N_c}{4I}\,\frac \partial {\partial x}\sum\limits_{-\omega _0<E_n\le
E_{lev}}\langle n|(1+\gamma ^0\gamma ^3)\tau ^3\gamma ^5\delta
(E_n+P^3-xM_N)|n\rangle \, .
\end{equation}
Although we use notations corresponding to the discrete spectrum actually
most of the spectrum is continuous. The singularities corresponding
appearing to $E_m=E_n$ are assumed to be regularized according to the
principal value prescription.

Making use of (\ref{du-minus-dd-occ}) we find
\begin{equation}
\left[ \Delta u(x)+\Delta d(x)\right] _{occ}^{\omega _0}=\left[ \Delta
u(x)+\Delta d(x)\right] _{occ}^{(1)\omega _0}-\frac 3{4IM_N}\,\frac \partial
{\partial x}\left[ \Delta u(x)-\Delta d(x)\right] _{occ}^{\omega _0}
\, , \label{Du-plus-Dd-occ-cutoff}
\end{equation}
where
\[
\left[ \Delta u(x)+\Delta d(x)\right] _{occ}^{(1)\omega _0}=\frac{N_cM_N}{2I%
}\sum\limits_m\sum\limits_{-\omega _0<E_n\le E_{lev}}\frac 1{E_n-E_m}
\]
\begin{equation}
\times \langle n|\tau ^3|m\rangle \langle m|(1+\gamma ^0\gamma ^3)\gamma
^5\delta (E_n+P^3-xM_N)|n\rangle \, .
\end{equation}
Similarly (\ref{Du-plus-Dd-non-occ}) leads to
\[
\left[ \Delta u(x)+\Delta d(x)\right] _{non-occ}^{\omega _0}=\left[ \Delta
u(x)+\Delta d(x)\right] _{non-occ}^{(1)\omega _0}
\]
\begin{equation}
-\frac 3{4IM_N}\,\frac \partial {\partial x}\left[ \Delta u(x)-\Delta d(x)%
\right] _{non-occ}^{\omega _0}  \, , \label{Du-plus-Dd-non-occ-cutoff}
\end{equation}
where
\[
\left[ \Delta u(x)+\Delta d(x)\right] _{non-occ}^{(1)\omega _0}=-\frac{N_cM_N}{%
2I}\sum\limits_m\sum\limits_{E_{lev}<E_n<\omega _0}\frac 1{E_n-E_m}
\]
\begin{equation}
\times \langle n|\tau ^3|m\rangle \langle m|(1+\gamma ^0\gamma ^3)\gamma
^5\delta (E_n+P^3-xM_N)|n\rangle \, .
\end{equation}
We see that
\[
\left[ \Delta u(x)+\Delta d(x)\right] _{occ}^{\omega _0}-\left[ \Delta
u(x)+\Delta d(x)\right] _{non-occ}^{\omega _0}
\]
\[
=\left[ \Delta u(x)+\Delta d(x)\right] _{occ}^{(1)\omega _0}-\left[ \Delta
u(x)+\Delta d(x)\right] _{non-occ}^{(1)\omega _0}
\]
\begin{equation}
-\frac 3{4IM_N}\,\frac \partial {\partial x}\left\{ \left[ \Delta
u(x)-\Delta d(x)\right] _{occ}^{\omega _0}-\left[ \Delta u(x)-\Delta d(x)%
\right] _{non-occ}^{\omega _0}\right\}  \, . \label{Du-Dd-anomaly-begin}
\end{equation}
Here
\[
\left[ \Delta u(x)+\Delta d(x)\right] _{occ}^{(1)\omega _0}-\left[ \Delta
u(x)+\Delta d(x)\right] _{non-occ}^{(1)\omega _0}
\]
\[
=\frac{N_cM_N}{2I}\sum\limits_m\sum\limits_{-\omega _0<E_n<\omega _0}\left(
\frac 1{E_n-E_m}\right) _{PV}
\]
\begin{equation}
\times \langle n|\tau ^3|m\rangle \langle m|(1+\gamma ^0\gamma ^3)\gamma
^5\delta (E_n+P^3-xM_N)|n\rangle \, .
\end{equation}
We remind that here the principal value prescription for $(E_n-E_m)^{-1}$ is
implied. This can be rewritten in the form

\[
\left[ \Delta u(x)+\Delta d(x)\right] _{occ}^{(1)\omega _0}-\left[ \Delta
u(x)+\Delta d(x)\right] _{non-occ}^{(1)\omega _0}=-\frac{M_NN_c}{4I}%
\int\limits_{-\omega _0}^{\omega _0}d\omega
\]
\[
\times \mathrm{Sp}\left\{ \left[ \left( \frac 1{H-\omega }\right)
_{P.V.}\tau ^3\delta (H-\omega )+\delta (H-\omega )\tau ^3\left( \frac
1{H-\omega }\right) _{P.V.}\right] \right.
\]
\[
\times \left. \delta (\omega +P^3-xM_N)(1+\gamma ^0\gamma ^3)\gamma
^5\right\} =-\frac{iM_NN_c}{4I}\int\limits_{-\omega _0}^{\omega _0}\frac{%
d\omega }{2\pi }
\]
\[
\times \mathrm{Sp}\left\{ \left[ \frac 1{H-\omega +i0}\tau ^3\frac
1{H-\omega +i0}\right] \delta (\omega +P^3-xM_N)(1+\gamma ^0\gamma ^3)\gamma
^5\right\}
\]
\[
+\frac{iM_NN_c}{4I}\int\limits_{-\omega _0}^{\omega _0}\frac{d\omega }{2\pi }
\]
\begin{equation}
\times \mathrm{Sp}\left\{ \left[ \frac 1{H-\omega -i0}\tau ^3\frac
1{H-\omega -i0}\right] \delta (\omega +P^3-xM_N)(1+\gamma ^0\gamma ^3)\gamma
^5\right\}  \, . \label{anomalous-difference-1}
\end{equation}

Hence
\[
\left[ \Delta u(x)+\Delta d(x)\right] _{occ}^{(1)\omega _{0}}-\left[ \Delta
u(x)+\Delta d(x)\right] _{non-occ}^{(1)\omega _{0}}
\]
\[
=-\mathrm{Im}\frac{M_{N}N_{c}}{2I}\int\limits_{-\omega _{0}}^{\omega _{0}}%
\frac{d\omega }{2\pi }\mathrm{Sp}\left\{ \frac{1}{H^{2}-\omega ^{2}-i0%
\mathrm{sign}\omega }(H+\omega )\tau ^{3}(H+\omega )\right.
\]
\[
\times \left. \frac{1}{H^{2}-\omega ^{2}-i0\mathrm{sign}\omega }\delta
(\omega +P^{3}-xM_{N})(1+\gamma ^{0}\gamma ^{3})\gamma ^{5}\right\}
\]
\[
=-\mathrm{Im}\frac{M_{N}N_{c}}{2I}\int\limits_{-\omega _{0}}^{\omega _{0}}%
\frac{d\omega }{2\pi }\mathrm{sign}\omega \mathrm{Sp}\left\{ \frac{1}{%
H^{2}-\omega ^{2}-i0}(H+\omega )\tau ^{3}(H+\omega )\right.
\]
\[
\times \left. \frac{1}{H^{2}-\omega ^{2}-i0}\delta (\omega
+P^{3}-xM_{N})(1+\gamma ^{0}\gamma ^{3})\gamma ^{5}\right\}
\]
\[
=-\mathrm{Im}\frac{M_{N}N_{c}}{2I}\int\limits_{-\omega _{0}}^{\omega _{0}}%
\frac{d\omega }{2\pi }\mathrm{sign}\omega \mathrm{Sp}\left\{ \frac{1}{%
-\partial ^{2}+M^{2}-\omega ^{2}-i0+iM(\gamma ^{k}\partial _{k}U^{\gamma
_{5}})}\right.
\]
\[
\times (\omega -i\gamma ^{0}\gamma ^{k}\partial _{k}+\gamma ^{0}MU^{\gamma
_{5}})\tau ^{3}(\omega -i\gamma ^{0}\gamma ^{k}\partial _{k}+\gamma
^{0}MU^{\gamma _{5}})
\]
\begin{equation}
\times \left. \frac{1}{-\partial ^{2}+M^{2}-\omega ^{2}-i0+iM(\gamma
^{k}\partial _{k}U^{\gamma _{5}})}\delta (\omega -i\partial
_{3}-xM_{N})(1+\gamma ^{0}\gamma ^{3})\gamma ^{5}\right\} \, .
\label{x-anomaly-start}
\end{equation}
The rest of the calculation is similar to how we worked with expression (\ref
{du-m-dd-start-0}) for the anomaly of $\Delta u(x)-\Delta d(x)$.

Nonzero contributions to the anomaly come from the expansion of the
propagators up to terms linear and quadratic in $iM(\gamma ^k\partial
_kU^{\gamma _5})$:
\[
\left[ \Delta u(x)+\Delta d(x)\right] _{occ}^{(1)\omega _0}-\left[ \Delta
u(x)+\Delta d(x)\right] _{non-occ}^{(1)\omega _0}
\]
\begin{equation}
=A_1(x)+A_2(x)  \, . \label{A1-A2-decomposition}
\end{equation}
Here $A_1(x)$ corresponds to terms linear in $iM(\gamma ^k\partial
_kU^{\gamma _5})$

\[
A_1(x)=\mathrm{Im}\frac{M_NN_c}{2I}\int\limits_{-\omega _0}^{\omega _0}\frac{%
d\omega }{2\pi }\mathrm{sign}\omega \mathrm{Sp}\left\{ \left[ iM(\gamma
^l\partial _lU^{\gamma _5})\frac 1{-\partial ^2+M^2-\omega ^2-i0}\right.
\right.
\]
\[
\times (\omega -i\gamma ^0\gamma ^k\partial _k+\gamma ^0MU^{\gamma _5})\tau
^3(\omega -i\gamma ^0\gamma ^k\partial _k+\gamma ^0MU^{\gamma _5})
\]
\[
+(\omega -i\gamma ^0\gamma ^k\partial _k+\gamma ^0MU^{\gamma _5})\tau
^3(\omega -i\gamma ^0\gamma ^k\partial _k+\gamma ^0MU^{\gamma _5})
\]
\[
\left. \times \frac 1{-\partial ^2+M^2-\omega ^2-i0}iM(\gamma ^l\partial
_lU^{\gamma _5})\right]
\]
\begin{equation}
\times \left. \left[ \frac 1{-\partial ^2+M^2-\omega ^2-i0}\right] ^2\delta
(\omega -i\partial _3-xM_N)(1+\gamma ^0\gamma ^3)\gamma ^5\right\} \, ,
\end{equation}
and $A_2$ is quadratic in $iM(\gamma ^k\partial _kU^{\gamma _5})$%
\[
A_2(x)=-\mathrm{Im}\frac{M_NN_c}{2I}\int\limits_{-\omega _0}^{\omega _0}%
\frac{d\omega }{2\pi }\mathrm{sign}\omega \mathrm{Tr}\left\{ \frac
1{-\partial ^2+M^2-\omega ^2-i0}\right.
\]
\[
\left[ iM(\gamma ^m\partial _mU^{\gamma _5})\frac 1{-\partial ^2+M^2-\omega
^2-i0}iM(\gamma ^n\partial _nU^{\gamma _5})\frac 1{-\partial ^2+M^2-\omega
^2-i0}\right.
\]
\[
\times (\omega -i\gamma ^0\gamma ^k\partial _k+\gamma ^0MU^{\gamma _5})\tau
^3(\omega -i\gamma ^0\gamma ^l\partial _l+\gamma ^0MU^{\gamma _5})
\]
\[
+iM(\gamma ^m\partial _mU^{\gamma _5})\frac 1{-\partial ^2+M^2-\omega ^2-i0}
\]
\[
\times (\omega -i\gamma ^0\gamma ^k\partial _k+\gamma ^0MU^{\gamma _5})\tau
^3(\omega -i\gamma ^0\gamma ^l\partial _l+\gamma ^0MU^{\gamma _5})
\]
\[
\times \frac 1{-\partial ^2+M^2-\omega ^2-i0}iM(\gamma ^n\partial
_nU^{\gamma _5})
\]
\[
+(\omega -i\gamma ^0\gamma ^k\partial _k+\gamma ^0MU^{\gamma _5})\tau
^3(\omega -i\gamma ^0\gamma ^l\partial _l+\gamma ^0MU^{\gamma _5})
\]
\[
\times \frac 1{-\partial ^2+M^2-\omega ^2-i0}iM(\gamma ^m\partial
_mU^{\gamma _5})
\]
\[
\left. \times \frac 1{-\partial ^2+M^2-\omega ^2-i0}iM(\gamma ^n\partial
_nU^{\gamma _5})\right]
\]
\begin{equation}
\times \left. \frac 1{-\partial ^2+M^2-\omega ^2-i0}\delta (\omega
-i\partial _3-xM_N)(1+\gamma ^0\gamma ^3)\gamma ^5\right\}  \, . \label{NLO-1}
\end{equation}
A straightforward calculation leads to the following results for $A_1(x)$
and $A_2(x)$%
\begin{equation}
A_1(x)=-\frac{M^2M_NN_c}{8\pi ^2I}\int \frac{d^3k}{(2\pi )^3}\frac 1{k^3}\ln
\left| 1+\frac{k^3}{xM_N}\right| \mathrm{Sp}\left\{ \tau ^3\left[ \tilde U(%
\mathbf{k)}\right] ^{+}\tilde U(\mathbf{k)}\right\}  \label{x-anomaly-res}
\end{equation}
\[
A_2(x)=\frac{M_NN_cM^2}{8\pi ^2I}\int \frac{d^3k}{(2\pi )^3}\ln \left| \frac{%
k^3+xM_N}{xM_N}\right|
\]
\begin{equation}
\times \left( \frac 1{k^3}+\frac 12\frac \partial {\partial k^3}\right)
\mathrm{Sp}\left\{ \tau ^3\left[ \tilde U(\mathbf{k)}\right] ^{+}\tilde U(%
\mathbf{k)}\right\} \, .
\end{equation}
Now we insert these results into (\ref{A1-A2-decomposition})
\[
\left[ \Delta u(x)+\Delta d(x)\right] _{occ}^{(1)\omega _0}-\left[ \Delta
u(x)+\Delta d(x)\right] _{non-occ}^{(1)\omega _0}
\]
\begin{equation}
=\frac{M_NN_cM^2}{16\pi ^2I}\int \frac{d^3k}{(2\pi )^3}\ln \left| \frac{%
k^3+xM_N}{xM_N}\right| \frac \partial {\partial k^3}\mathrm{Sp}\left\{ \tau
^3\left[ \tilde U(\mathbf{k)}\right] ^{+}\tilde U(\mathbf{k)}\right\} \, .
\end{equation}
Note that shifting the integration variable
\begin{equation}
k^3\rightarrow k^3-xM_N \, ,
\end{equation}
we obtain
\[
\int \frac{d^3k}{(2\pi )^3}\ln \left| \frac{k^3+xM_N}{xM_N}\right| \frac
\partial {\partial k^3}\mathrm{Sp}\left\{ \tau ^3\left[ \tilde U(\mathbf{k)}%
\right] ^{+}\tilde U(\mathbf{k)}\right\}
\]
\begin{equation}
=-\frac 1{M_N}\frac \partial {\partial x}\int \frac{d^3k}{(2\pi )^3}\ln
\left| k^3+xM_N\right| \mathrm{Sp}\left\{ \tau ^3\left[ \tilde U(\mathbf{k)}%
\right] ^{+}\tilde U(\mathbf{k)}\right\} \, .
\end{equation}

Therefore
\[
\left[ \Delta u(x)+\Delta d(x)\right] _{occ}^{(1)\omega _0}-\left[ \Delta
u(x)+\Delta d(x)\right] _{non-occ}^{(1)\omega _0}
\]
\begin{equation}
=-\frac{N_cM^2}{16\pi ^2I}\frac \partial {\partial x}\int \frac{d^3k}{(2\pi
)^3}\ln \left| \frac{k^3+xM_N}{xM_N}\right| \mathrm{Sp}\left\{ \tau ^3\left[
\tilde U(\mathbf{k)}\right] ^{+}\tilde U(\mathbf{k)}\right\} \, .
\end{equation}
Inserting this result and (\ref{du-minus-dd-anomaly-res}) into (\ref
{Du-Dd-anomaly-begin}) we observe a complete cancellation:
\begin{equation}
\left[ \Delta u(x)+\Delta d(x)\right] _{occ}^{\omega _0}-\left[ \Delta
u(x)+\Delta d(x)\right] _{non-occ}^{\omega _0}=0
\end{equation}

Thus the isoscalar polarized quark distribution $\Delta u(x)+\Delta d(x)$ is
nonanomalous.

Using similar methods one can check that function $\Delta u(x)+\Delta d(x)$
is free of ultraviolet divergences: although the two separate terms in the
rhs of (\ref{Du-plus-Dd-occ-cutoff}) are UV divergent the total sum is
finite.

\section{Numerical results}

\label{num-res-section}

The numerical results for the isovector polarized distribution function $%
\Delta u(x)-\Delta d(x)$ are given in \cite{DPPPW97}. For the computation of $%
\Delta u(x)+\Delta d(x)$ (\ref{Du-plus-Dd-occ}), (\ref{Du-plus-Dd-non-occ}) we use
the numerical methods which were developed in \cite{DPPPW97} and later
extended in  \cite{PPGWW} for the computation of the isovector unpolarized
distribution.

The eigenvectors and eigenvalues of the Dirac Hamiltonian (\ref{H-Dirac})
are determined by diagonalizing in the free Hamiltonian basis (\ref{freeham}).
This basis is made discrete by placing the soliton in a
three-dimensional
spherical box of finite radius $D$ and imposing the Kahana-Ripka boundary
conditions \cite{KR84}. Both $\Delta u(x)-\Delta d(x)$ and $\Delta u(x)+\Delta d(x)$
were computed using the standard value of the constituent quark mass $M=350$
MeV as derived from the instanton vacuum \cite{DP86}.

In our calculation we use the self-consistent solitonic profile $P(r)$ (see
e.g. ref. \cite{Doering92,WG97}). However, performing the numerical
calculations in the finite spherical box one should be careful about the
large distance effects. To be safe, we artificially exponentially suppress
the pion tail of the soliton profile at large distances so that the field
vanishes outside the box (a similar problem in the calculation of $g_{A}$
was studied in \cite{g-A-massless}).

\begin{figure}[tbp]
\epsfxsize=6cm 
\centerline{\epsffile{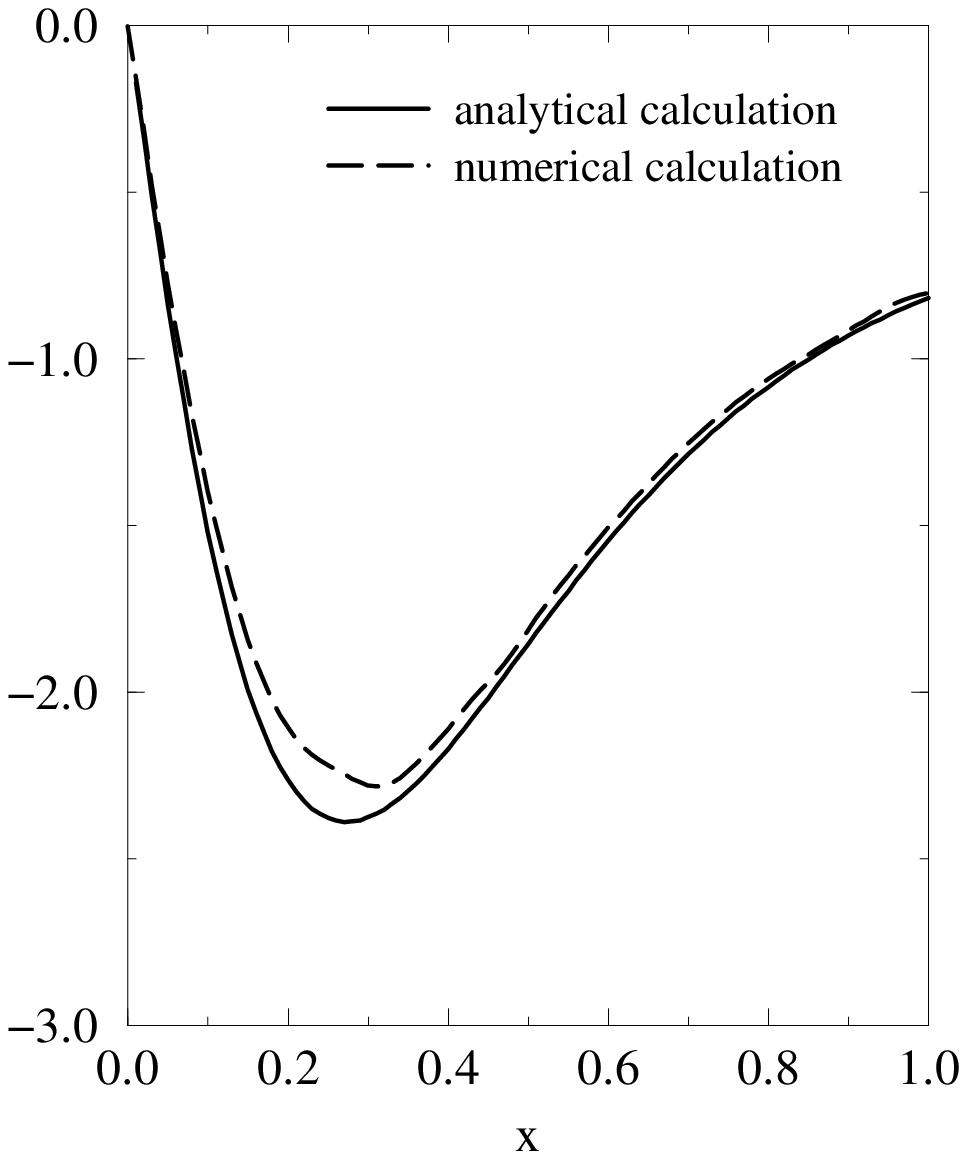}}
\caption{Analytical (solid) and numerical (dashed) results for the anomalous
difference $[\Delta u-\Delta d]_{occ}-[\Delta u-\Delta d]_{non-occ}$.}
\label{isovector-polarized-anomaly}
\end{figure}

In Fig. \ref{isovector-polarized-anomaly} we compare our numerical results
for the anomaly of $\Delta u(x)-\Delta d(x)$ with the analytical result (\ref{du-minus-dd-anomaly-res}).
We observe a rather good agreement.

\begin{figure}[tbp]
\epsfxsize=6cm 
\centerline{\epsffile{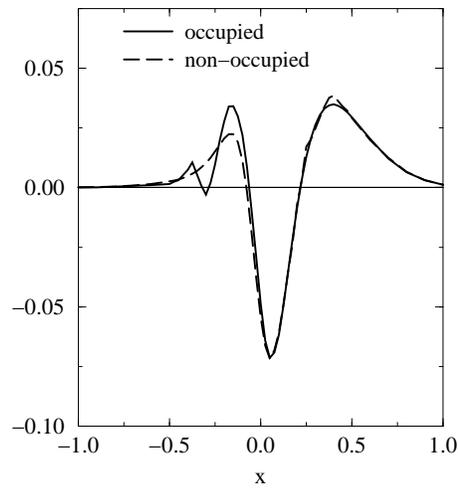}}
\caption{Results for continuum contribution $[\Delta u+\Delta d]_{sea}$ based
on the occupied and non-occupied representations.}
\label{isosinglet-polarized-anomaly}
\end{figure}

Fig. \ref{isosinglet-polarized-anomaly} shows the numerical results for the
Dirac sea contribution to $\Delta u(x)+\Delta d(x)$ based on the two
representations (occupied and non-occupied). We see a reasonable agreement
between the two results which confirms the absence of the anomaly in $\Delta
u(x)+\Delta d(x)$. Some difference between the two curves at negative $x$ is
finite-box artefact. Increasing the size of the box one can see that this
difference tends to disappear.

In Fig. 3 we compare the result of the calculation of $\Delta u(x)+\Delta d(x)$, $
\Delta \bar{u}(x)+\Delta \bar{d}(x)$ with the GRSV-LO parametrization \cite{GRSV-pol}
at the low scale of the model $\mu= 600 \, {\rm MeV}$.
We see that the quark distribution $\Delta u(x)+\Delta d(x)$
is in a reasonable
agreement with the GRSV parametrization whereas the antiquark distribution $%
\Delta \bar{u}(x)+\Delta \bar{d}(x)$ obtained in the model is considerably smaller
than that of the GRSV parametrization.
Note that the polarized antiquark distributions are not
directly accessible in inclusive hard reactions. Due to
the lack of data the GRSV parametrizations therefore are
based on certain assumptions, e.g. in the GRSV analysis
it was assumed that $\Delta \bar{u}(x)=\Delta \bar{d}(x)$.
In contrast to this the QCD large $N_{c}$ counting and the
quark soliton model predict a large flavour asymmetry
$\Delta \bar{u}(x) > \Delta \bar{d}(x)$ in the light
polarized sea. Some physical applications of this have been studied in
refs.~\cite{WW-99,semi,dy}.

Fig. 4 shows our predictions for the polarized antiquark distributions
$\Delta \bar{u}(x)$ and $\Delta \bar{d}(x)$ separately at the scale
$\mu = 600\,{\rm MeV}$.

Since the quark distribution $\Delta u+\Delta d$ is finite, no ultraviolet
regularization is needed for this quantity. There is even an
argument against
regularizing $\Delta u+\Delta d$ coming from the fact that the first moment
of this distribution is related to the imaginary part of the quark
determinant in the background soliton field which has to be left
nonregularized if one wants to keep baryon number conserved -- this is an
analog of the nonrenormalizability of the Wess-Zumino term in pure chiral
models.

Several comments should be made about the calculations of $\Delta u+\Delta d$
within the same model by Wakamatsu et al. who published three different
versions of the calculation in papers \cite{WK-97, WK99, WW-99}. In paper
\cite{WK-97} one of the terms was overlooked. This mistake was corrected by
the authors of \cite{PPGWW}. The revised version of calculation of Wakamatsu
et al. was published in \cite{WK99}. In this paper the question about the
anomalous difference $\left[ \Delta u(x)+\Delta d(x)\right] _{occ}^{\omega
_{0}}-\left[ \Delta u(x)+\Delta d(x)\right] _{non-occ}^{\omega _{0}}$ was
investigated only numerically but the accuracy of the calculation did not
allow the authors to draw any conclusions concerning whether this difference
vanishes or not. Actually the numerical accuracy of the agreement between
the two representations which we observe in our calculation (see Fig. \ref
{isosinglet-polarized-anomaly}), and which is necessary for a proper evaluation of the parton
distributions, is of two orders magnitude better than the same
difference presented in \cite{WK99}. The practical solution accepted in \cite
{WK99} was to use $\left[ \Delta u(x)+\Delta d(x)\right] _{occ}^{\omega
_{0}} $ for $x>0$ and $\left[ \Delta u(x)+\Delta d(x)\right]
_{non-occ}^{\omega _{0}}$ for $x<0$ (i.e. for the antiquark distribution).
As it was explained above, $\Delta u(x)+\Delta d(x)$ should not be
regularized contrary to what the authors of \cite{WK99} do.


The polarized structure functions were also estimated in the work
\cite{WGR-97} in the Nambu--Jona-Lasinio model within the
valence approximation. In ref.~\cite{DPPPW96} it was shown that the
valence approximation leads to a number of inconsistencies:
antiquark distributions are negative, sum rules are violated, etc.

Our work has been performed within the quark-soliton model with two quark
flavors. In the case of the model in the flavor $SU(3)$  the same quantity
should be interpreted as $\Delta u+\Delta d+\Delta s$.

The first moment of the $\Delta u(x)+\Delta d(x)$ gives the singlet axial
charge. Our  result of $g^{(0)}_A = \int^{1}_{-1} dx (\Delta u+\Delta d)(x) = 0.35$
agrees with the calculation performed in other works \cite
{WY91,BPG93}. Note that in the calculation of this charge no ultraviolet
regularization was used.

\section{Conclusions}

We have proved that the representation of singlet polarized (anti)quark
distributions in the chiral quark-soliton model as a sum over quark orbitals
is ultraviolet finite and free of quantum anomalies. This is a serious check
of the consistency of the quark-soliton model.

In fact, the cancellation of quantum anomalies in the model is related to the
fact that certain basic properties of QCD as a local quantum field theory
are realized in the model. The equivalence of the summation over
occupied and non-occupied states is directly connected to anticommutativity
of fermion fields at space-like intervals. Actually this locality property
has direct relation to the positivity of quark and antiquark densities in
the quark soliton model \cite{DPPPW96,DPPPW97}.

Another consequence of the cancellation of anomalies is that the model
results for the
parton distributions are compatible with the charge conjugation invariance:
the quark distributions in nucleon coincides with the antiquark distributions
in the antinucleon.

{}From the practical point of view the results presented in this paper
allow us to conclude that for the calculation of the singlet polarized
quark and antiquark
distributions no Pauli-Villars subtraction is needed. Additionally the
numerical check of the cancellation of the anomalies is a powerful tool to
control the accuracy of the numerics.

We have computed the singlet polarized quark and antiquark distributions which
arise in the subleading order of $1/N_c$ expansion. We found the quark distribution $%
\Delta u(x)+\Delta d(x)$  to be in a reasonable agreement with GRSV
\cite{GRSV-pol} parametrization of parton distributions at low normalization
point. A remarkable prediction of our model is that the polarized
distributions of $u$ and $d$ antiquarks are essentially different, see
Fig.4. Usually, in parametrizations of polarized parton distributions, it
was assumed that $\Delta \bar u(x)=\Delta \bar d(x)$, which is not confirmed
by our model calculations (see Fig.~4). It would be extremely interesting to
include into the fits of the data the flavour decomposition pattern for
polarized antiquarks obtained in our model calculations.
Future experiments at HERA and RHIC investigating Drell-Yan
lepton pair production in polarized nucleon-nucleon collisions
will clarify the situation. For a discussion see \cite{semi,dy}.
Let us note that in the singlet polarized channel under the evolution the
quark distributions mix with polarized gluon distribution. Analysis of ref.~
\cite{DPW96,DPPPW96} in the framework of the instanton model of the QCD
vacuum shows that the gluon distribution is parametrically smaller
(suppressed by $M^{2}/M_{PV}^{2}$) than quark and antiquark distributions.
In order to obtain a non-zero result one has to go beyond the zero-mode
approximation of ref. \cite{DPW96} and/or consider contributions of many
instantons. Both ways would lead to extra powers of the packing fraction of
instantons. This means that gluons at low normalization point inside the
nucleons appear only at the level of $M^{2}/M_{PV}^{2}$.

\vspace{1cm}

\noindent{{\bf Acknowledgments}}

We are grateful to N.-Y. Lee, V.Yu. Petrov, T. Watabe and
C. Weiss for numerous interesting discussions.

One of the authors (P.V.P.) is grateful to the Institute for Theoretical
Physics II of the Ruhr University Bochum for warm hospitality.
P.V.P. and M.V.P. have been partially
supported by RFBR grant 96-15-96764.
D.U. acknowledges the financial support from PRAXIS XXI/BD/9300/96 and
from PRAXIS PCEX/C/FIS/6/96. The work has partialy been supported by
DFG and BMFB.

\newpage

\begin{figure}[tbp]
\epsfxsize=6cm 
\centerline{{\epsffile{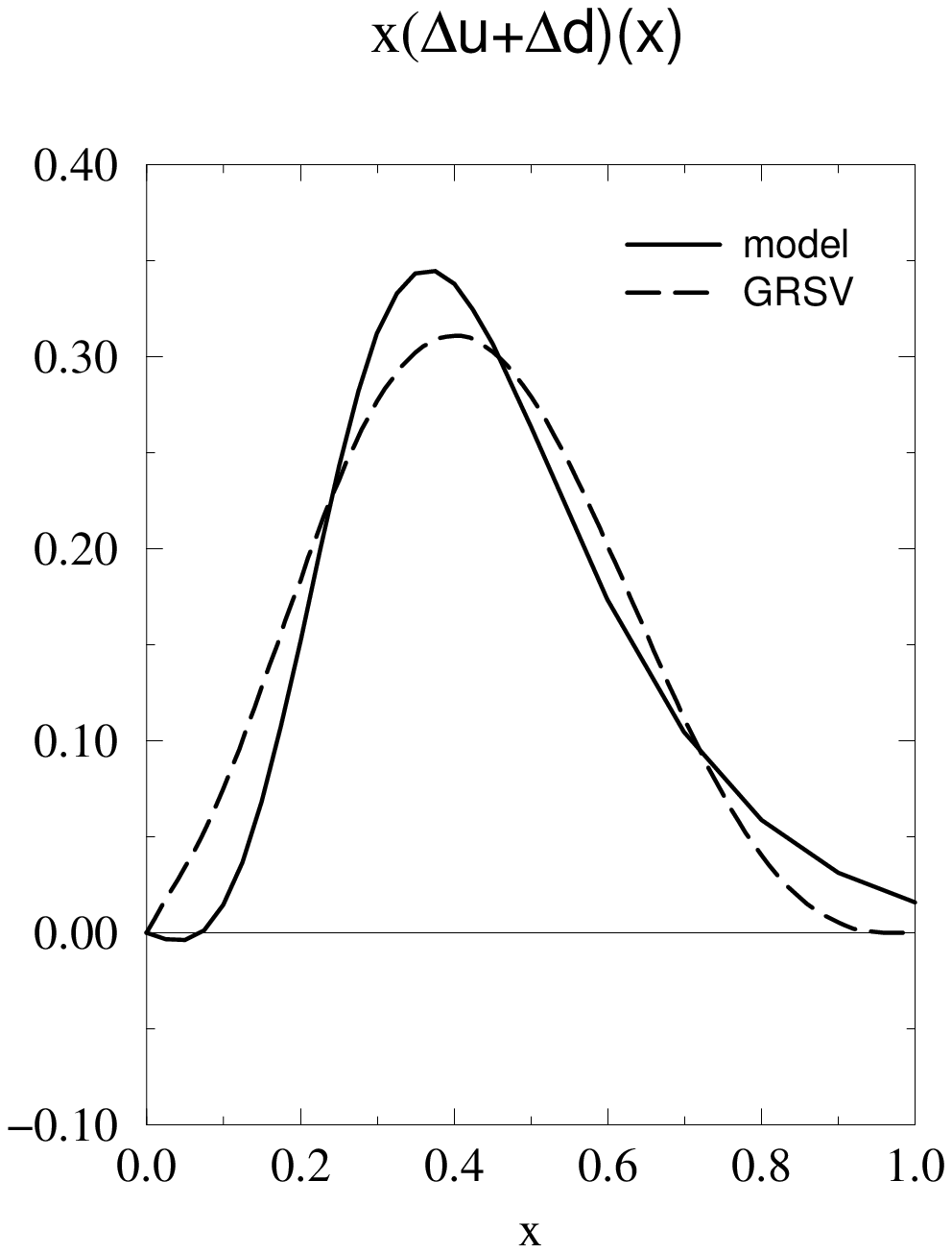}}{\hspace*{1cm}}
{\epsfxsize=6cm\epsffile{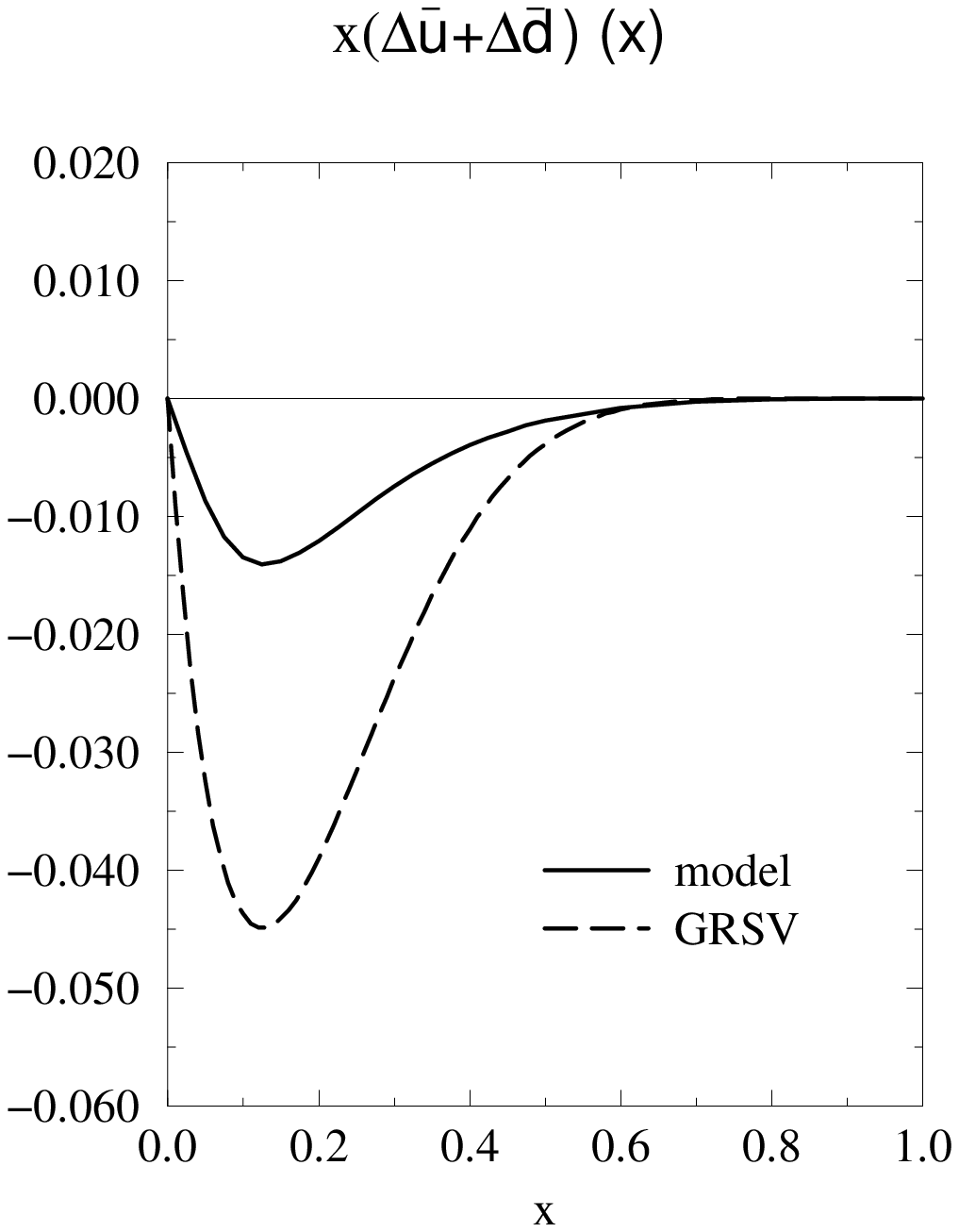}}}
\caption{The quark soliton model results for $x[\Delta u+\Delta d]$ and $%
x[\Delta{\bar u}+\Delta{\bar d}]$ versus LO-GRSV parametrization at the scale
$\mu \approx 600MeV$.}
\label{Delta-u-d-plot}
\end{figure}

\begin{figure}[tbp]
\epsfxsize=6cm 
\centerline{\epsffile{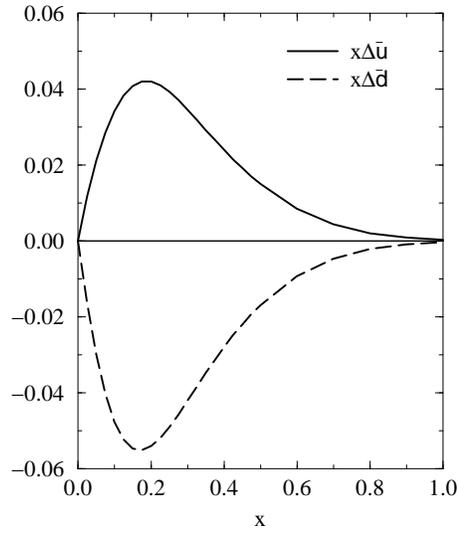}}
\caption{The quark soliton model predictions for $x\Delta{\bar u}$ and  $x\Delta{\bar d}$
at the scale $\mu \approx 600MeV$.}
\label{Delta-antiu-antid-plot}
\end{figure}

\end{document}